\documentclass{llncs}
\usepackage{graphicx}
\usepackage{pifont}
\usepackage{algorithm}
\usepackage{algorithmic}

\newcommand{\tick}{\ding{52}}
\newcommand{\cross}{\ding{56}} 
\renewcommand{\algorithmicrequire}{\textbf{Input:}}
\renewcommand{\algorithmicensure}{\textbf{Output:}}

\begin{document}
%
%
\title{Using Image Attributes for Human Identification Protocols}
\author{Hassan Jameel\inst{1} \and Heejo Lee\inst{2} \and Sungyoung Lee\inst{1}}


\institute{Department of Computer Engineering, Kyung Hee University,
449-701 Suwon, South Korea\\
\email{\{hassan,sylee\}@oslab.khu.ac.kr},\\ 
\and
Department of Computer Science and Engineering, Korea University
Anam-dong, Seongbuk-gu, Seoul 136-701, South Korea\\
\email{heejo@korea.ac.kr}}
\maketitle
\begin{abstract}
A secure human identification protocol aims at authenticating human users to a remote server when even the users' inputs are not hidden from an adversary. Recently, the authors proposed a human identification protocol in the RSA Conference 2007, which is loosely based on the ability of humans to efficiently process an image. The advantage being that an automated adversary is not effective in attacking the protocol without human assistance. This paper extends that work by trying to solve some of the open problems. First, we analyze the complexity of defeating the proposed protocols by quantifying the workload of a human adversary. Secondly, we propose a new construction based on textual CAPTCHAs (Reverse Turing Tests) in order to make the generation of automated challenges easier. We also present a brief experiment involving real human users to find out the number of possible attributes in a given image and give some guidelines for the selection of challenge questions based on the results. Finally, we analyze the previously proposed protocol in detail for the relationship between the secrets. Our results show that we can construct human identification protocols based on image evaluation with reasonably ``quantified'' security guarantees based on our model.
\end{abstract}
\section{Introduction}
Suppose a student wishes to write a confidential email to disclose the information of a leaked out exam paper to his friend.  Using a secure email client the student writes down an email, sends it and logs out. The email shall be encrypted and would only be viewable by the recipient once he logs in to check the email. How can the student be sure that the email was sent securely and no one could learn anything apart from   the intended recipient? The answer relies on the weakest security link: the password. Little did the student know that his computer was being key-logged \cite{keylog}. There was a hidden camera looking at the student's every move. His fellow student also shoulder surfed on the password. Even if no one saw the mail being written, they could log on later to view the sent mail box. It turns out that no matter how secure the email client was, it only served its purpose until the password was not compromised.

Similar situation occurs when one inputs PIN numbers on ATMs. We could use biometrics instead of passwords or pin numbers. But biometric data is only secure unless the biometric information is kept confidential and the equipment has not been tampered. So, are these mechanisms unsecure or useless? They certainly aren't. They were designed with certain assumptions in mind: The passwords selected by the user should be truly random strings of a suitable length; the pin numbers selected by the user should be truly random 4-digit numbers; the user has the responsibility to hide her input from peeping eyes. It has been an interesting topic of research in cryptography to devise authentication protocols that meet the security reequirements even when the above mentioned assumptions do not hold. This highly vulnerable security  environment has been termed as the ``naked human in a glass house'' model in \cite{humanoid}, although the first protocol constructed to be secure in this model was by Matsumoto \cite{mat:ima}. We shall call such protocols as ``Human Identification Protocols'' or HIPs in short following the terminology in literature. A lot of human identification protocols have been proposed in literature with the goal of ease of ``human execution'' in mind. The protocols should take roughly the same amount of time as a password based protocol takes. Researchers have tried to construct protocols that are secure and require little or no computation on the human's part. This indeed is a very hard goal to achieve and reflects through the fact that there are not a lot of proposed human identification protocols over the years.

The situation, however, is not as bad as it seems. Humans possess good cognitive abilities. We can recall a previously viewed image with a very high probability when presented to us again. We do not need to memorise all the details of the picture. This has led to the use of graphics as essential ingredients of human identification protocols. Proposed identification protocols can be implemented with a graphical implementation. For example, instead of memorizing the string 00101, we can display five pictures, each time, where the third and the fifth picture is shown to the user apriori as the user's secret pictures. We could go a step further and use the things that an image describes to build human identification protocols. In \cite{hassan} we proposed that instead of using pictures just as memory aids, we can use their internal structure in some way to construct a human identification protocol. The secret could be one of the concepts that the picture satisfies. It was conjectured that it would be hard even for a human adversary to find the secret. However, no exact quantification of this hardness was given. While the hardness of breaking the protocol is clearly evident against automated adversaries (computer programs programmed to defeat the protocol), the security against ``human adversaries'' needs more attention. This study aims to enhance the work proposed in \cite{hassan} by answering some of the open problems and moves a step further in trying to quantify the hardness of the underlying problem. We address the following issues:
\begin{itemize}
\item We present a new protocol with the aim of generating automated instances of challenges. This construction is loosely based on the Gimpy CAPTCHAs \cite{gimpy} and requires the server to only maintain a dictionary of words.
\item We analyze the security of the protocol presented in \cite{hassan} and the ones presented in this paper with a new perspective. We show how much work has to be performed by a ``human'' adversary in order to obtain the secret based on our model.
\item We show the interrelationship of the two ``secrets'' of the protocol presented in \cite{hassan}.
\item We also present another way of viewing the underlying problem of these protocols using matrix representation. This view helps to understand the principal hardness of the protocols presented.
\item Finally, we show the results of some experiments which show the amount of information in a very simple image. This data leads us to some guidelines while selecting the secret which we have also mentioned.
\end{itemize}
Our study will deal with passive adversaries only. The reason being that if we deal with active adversaries in the protocols, then we might require the human user to send some random challenges to the remote server in order to authenticate it as well. This generation of random challenges requires an extra amount of computation from the human user's part, and might deem the protocols impractical. This is left as a future work.    
\section{Related Work}
The first work on human identification dates back to \cite{mat:ima}. Since then a lot of other schemes have been proposed in literature \cite{wang},\cite{mat},\cite{xiang},\cite{hopper},\cite{shuj},\cite{wein}. Some of them were broken in \cite{wang}, \cite{goll}. While most of them involve some numerical calculations like \cite{mat:ima} and the HB protocol\cite{hopper}, they can be implemented using some graphical interface employing pictures as memory aids. We can categorized the human identification protocols into two broad categories: Protocols built to be secure against general eavesdropping adversaries and protocols secure against only ``guessing adversaries'' i.e. Adversaries who do not see the user's input and hence try to guess the secret or impersonate the user without any apriori knowledge. Protocols mentioned so far fall in the first category. They have a drawback, however, that they involve extra computation from the user. As an example, in the HB protocol \cite{hopper}, the user is required to compute bit-wise binary multiplication for some number of bits in every iteration. This may not seem much but to obtain a higher level of security, the number of computations increase significantly.

In the second category, the most well known example is the traditional password based authentication system. Others include purely graphical schemes like DeJa Vu\cite{dham},Passface \cite{passf}, Point \& Click \cite{passp} and \cite{grap} that require little or no numerical computation whatsoever. The basic theme of \cite{dham} and \cite{passf} is to present the user a series of pictures, a subset of which are the secret pictures. The user is authenticated if its selection of the secret pictures among the given set of pictures is correct. On the other hand, in \cite{passp} the user is authenticated if it clicks on the correct secret location in the given picture. \cite{grap} works similarly by letting the user draw the secret symbol or figure on a display device. Evidently, these purely graphical schemes are not secure against ``peeping'' attacks \cite{shuj}. Anyone observing the actions of the user can find out the secret in no time. For a detailed account of all the schemes, see \cite{shuj}. In our previous work \cite{hassan}, we proposed to use internal properties of images as secrets. After a secret has been chosen, pictures which satisfy the properties were presented randomly with the pictures that do not satisfy the property. The user has to answer the pictures according to the secret property. It was conjectured that finding out the secret property is a hard problem for adversaries. For automated adversaries, this follows immediately from the definition of CAPTCHAs\cite{luis}. But for human adversaries, this hardness is difficult to prove. In this paper, we have tried to quantify this hardness and tried to answer some of the open questions described above. 
\section{The Main Idea}
Consider the picture of a magic square shown in Figure 1. How many things does the picture represent? A simple glance at it can reveal a lot of things: a magic square, a square, nine small squares, digits, black, white, the digit 4, the digit 2, line(s), the right angle etc. It is amazing how much information does a rather simple looking picture contain. We call each piece of this information as a feature. We can take one of these features and construct a question out of it. For example: Does the picture contain a rectangle?. This question is shared as a secret between the server and the user. After that, a series of pictures can be presented to the user such that with probability 1/2 they satisfy the question and with probability 1/2 they don't. The user has to scan the picture and answer `yes' or `no' accordingly. How about the adversary? The adversary would like to know the secret feature. The best way to do it is to extract all features in the pictures presented to the user and then do intersection (if user's answer is `yes') or difference (if user's answer is `no') to narrow down the number of possible secret features. 
\begin{figure}[ht]
\centerline{\includegraphics[width=3.63in,height=2.87in]{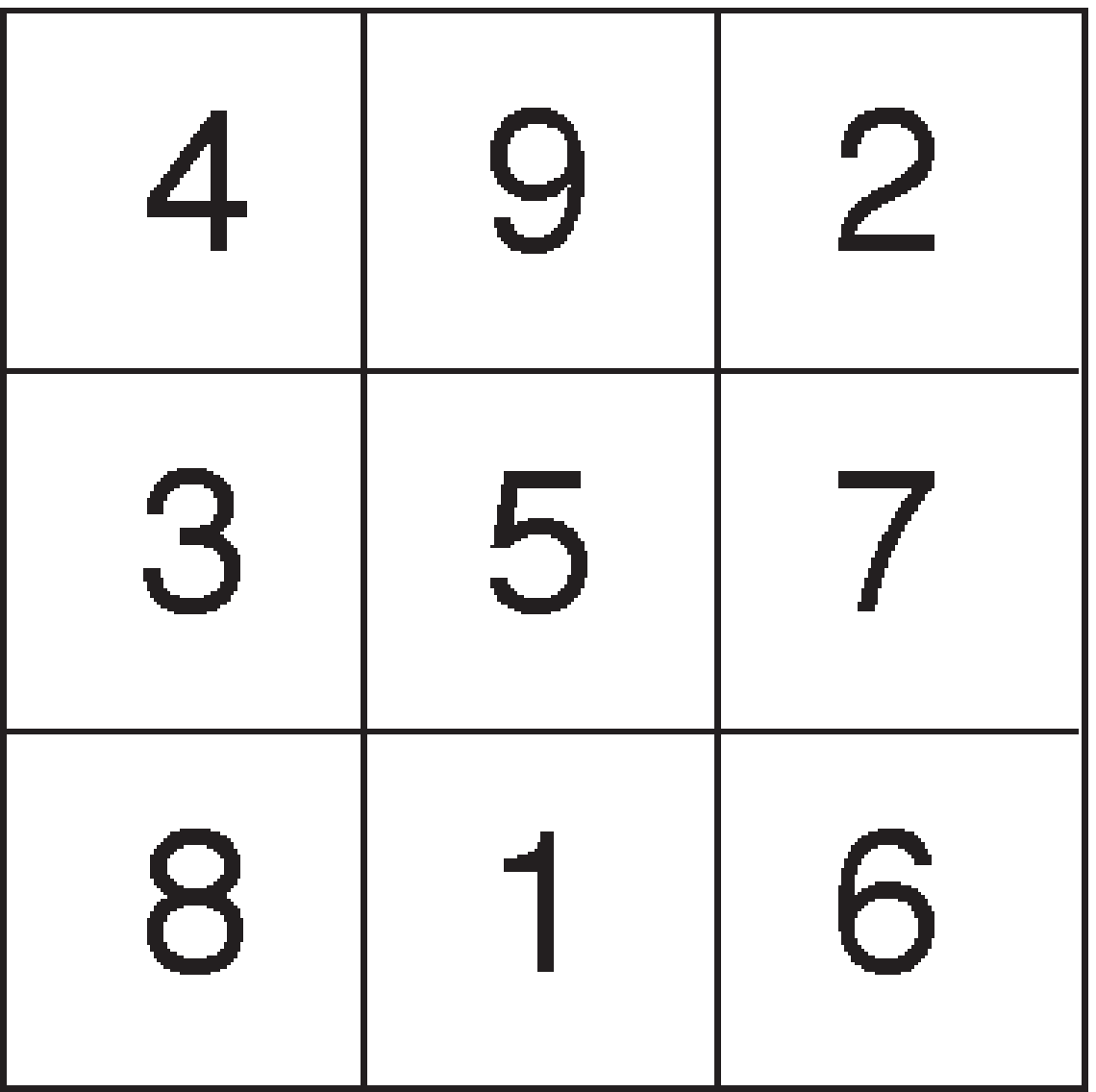}}
\caption{A simple picture of a magic square may describe a lot of things}
\label{fig5}
\end{figure}

From an abstract point of view, we may define a universal set of all features, all pictures could possibly have. Denote this set as $S = \left\{ {1,2, \ldots ,n} \right\}$, where $n$ is the total number of features. Any subset of this set represents a picture. Given a subset $A$ of this set (which again is a picture) and its corresponding response bit $a$ from the user, we would be interested in finding out how the adversary can find the hidden feature and hence the secret question. Obviously, it is almost impossible to write a computer program which given a picture can filter out all the features that picture describes considering the enormous size of the set of all features. This immediately implies the need of human intervention. So, what could be the best strategy to find out the hidden feature? One way is to have a cursory glance at the pictures and see if there is something common between the pictures. A more efficient way is to find out all the features in the pictures, and then check which features are common between the pictures. But once the features have been extracted, one could make a program that would automatically filter out the common features. Thus we will only be concerned with the workload on the human adversary which amounts to finding out all the features in the pictures until a single common feature is found. Our analysis will try to quantify the complexity of our proposed schemes based on the workload on human adversary using the mentioned approach. Namely, given a picture $A$ and its corresponding answer bit $a$ the adversary has to flag all the features as the candidate secret features if $a=1$ or else delete all the features present in this picture from the candidate set of features if $a=0$. Thus the adversary's job is to find out the features by personally checking every picture and narrow down the set of candidate secret features.\\
We will now present protocols based on this main idea in the next section. Each protocol follows a short discussion and a brief account of the workload on the adversary. The detailed analysis follows in Section VI and VII.

\section{Identification Schemes}
Before we present the proposed schemes we present the general notation which will be used throughout this manuscript. We assume there to be a pool of distinct pictures ${\cal P}$, each element of which is denoted by $P$, and a set of questions ${\cal Q}$. Each question $q$ has a binary answer when applied to any picture in ${\cal P}$. Each question therefore asks whether a certain feature is present in the picture or not. We will also use $q$ to represent the function: $q:{\cal P} \to \left\{ {0,1} \right\}$, which represents the evaluation of a picture according to the question. The user's answer string is represented by $A$, where $A\left( i \right)$ represents the $i$th bit in the string. We will use $a$ to denote an arbitrary answer bit. From now onwards, the word ``adversary'' or the symbol ${\cal H}$ would mean the ``human'' eavesdropping adversary, unless otherwise specified. The workload of ${\cal H}$ for each protocol will be based on the complexity of the above mentioned algorithm (or a slight variant of it) which is described in Section VI. Due to a bulk of notation used in this article, we will abuse this general notation in some sections or subsections without compromising disambiguity.
\subsection{The Basic Scheme}
We have the following immediate basic scheme:\\
\textsc{Setup.} The user and the server share a secret question $q$ from ${\cal Q}$.\\
\textsc{Protocol.}
\begin{itemize}
\item Repeat $k$ times
	\begin{itemize}
	\item The server picks a bit $b$ uniformly at random, and picks a picture $P$ from ${\cal P}$ such that $q\left( P \right) = b$. Discards this picture from the pool, and presents it to the user.
	\item The user submits $a = q\left( P \right)$.
	\end{itemize}
\item Output $\mathsf{accept}$ if all answer's are correct, otherwise output $\mathsf{reject}$
\end{itemize}

The scheme is described pictorically in Figure 2, where $2$ pictures are shown at each iteration. 
\begin{figure}[ht]
\centerline{\includegraphics[width=3.63in,height=2.87in]{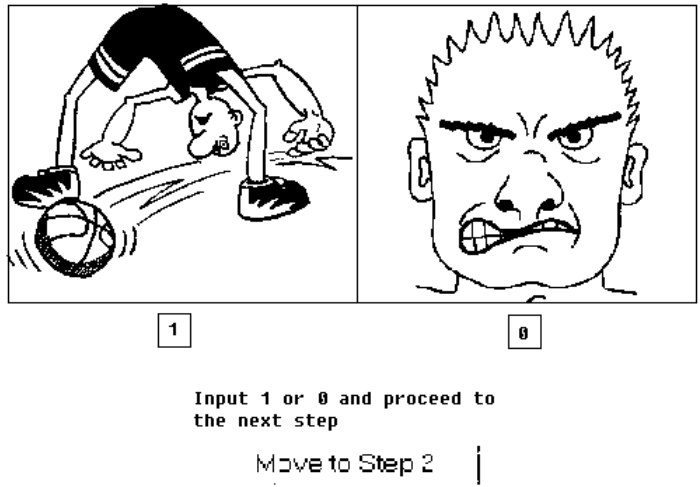}}
\caption{The Basic Protocol with $L=2$ and secret question $q=$``Does the picture contain a basketball?''}
\label{fig6}
\end{figure}
We could present all $k$ pictures at the same time depending on whether they can be displayed on the screen or not. In anycase we can find a number $L$ such that we can have $\left\lceil {{k \mathord{\left/
 {\vphantom {k L}} \right.
 \kern-\nulldelimiterspace} L}} \right\rceil $
iterations in one authentication session. According to the analysis in Section VI, the number of pictures the adversary has to observe to obtain the secret feature would be: $\log _2 n$ regardless of the value of $L$. This amounts to a total work of $\frac{n}{2}\log _2 n$, which has to be done by the human adversary. Setting in the values $n = 10^6$, we get a total work of $ \approx 2^{23} $ units.
\paragraph*{Discussion on the protocol} This protocol is simple and practical. However, there is a big disadvantage. Namely, the user is sending its answers in the clear. An adversary thus knows the ``correct'' answer to all the pictures shown to the user. This makes the life of the adversary a bit easier since it can have a glance at the pictures to find out the common secret feature. We would like to somehow ``hide'' the answer sequence, thus making it hard for the adversary to guess the secret feature. The next protocol attempts to do that. This protocol was the original protocol presented in \cite{hassan}.
\paragraph*{Matrix Interpretation}Consider the set of all features $S = \left\{ {1,2, \ldots ,n} \right\}$ as described in the previous section. Each picture contains a subset of this universal set of features. The secret question can also be considered as a subset of this set with a single element. We can represent the features of a presented picture $i$ as a vector ${\bf{v}}_i$ with all the features not present in the picture represented by 0's, and the secret question as ${\bf{x}}$ with only one entry equal to 1. We can then write the answer bit as ${\bf{v}}_i  \cdot {\bf{x}} = a$. Thus if we have more pictures, we can represent the protocol as the matrix operation: ${\bf{Vx}} = {\bf{a}}$, where ${\bf{V}}$ is the matrix containing row vectors $v_i$ representing the features of the picture $i$, and ${\bf{a}}$ is the answer vector. The obvious way to solve this requires $O\left( n \right)$ pictures and corresponding answers. However, as we will see in Section VI, not all pictures will contain features from the full feature space. Therefore, the actual number of picture and answer pairs required would be less than that. The real difficulty of the problem, however, is to find out the features in the pictures and thus construct the correct matrix ${\bf{V}}$. Our matrix representation shows one way of describing the problem needed to be solved by ${\cal H}$.
\subsection{The Enhanced Protocol} 
Suppose we want to present $L$ pictures at a time to the user. The pictures are labeled from $1$ to $L$ in sequential order. The idea is to permute the numbers randomly. Out of the resulting permutation, select $l$ numbers, and label the others as ``don't care positions''. This permutation string is also shared as a secret between the user and the server. Each time a series of $L$ pictures are presented. The user answer's the pictures according to the order in the permutation string, and fills the ``don't care positions'' with random bits. So, for example, if $L=10$ and $l=5$, then a possible permutation string would be $**54*39**0$, where the ``$*$'' represents a don't care position. We label such a permutation string as $\sigma $. We now have two secrets in our scheme: the secret question $q$ and the permutation string $\sigma $. The user thus has to answer a series of pictures according to the function: $q_\sigma  :{\cal P} \to \left\{ {0,1} \right\}$. The user is accepted if the answers are correct at the $l$ positions in each of the $k$ iterations. The rest of the protocol is the same as the basic protocol.

The protocol is described in Figure 3. Here we have taken $L=5$, $l=3$ and $k=1$. The permutation string is $2*43*$ and the secret question is  ``Is the picture somehow related to computation?''. 
\begin{figure}[ht]
\centerline{\includegraphics[width=4.63in,height=2.87in]{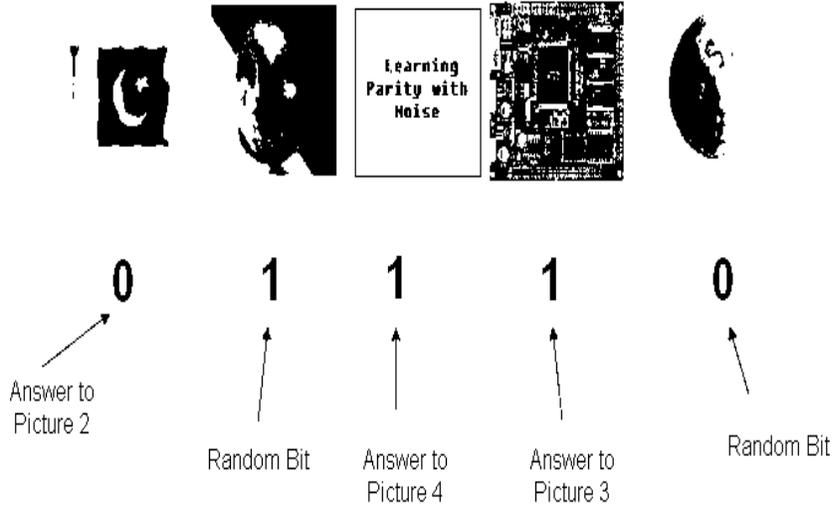}}
\caption{The Enhanced Protocol with $L=5$, $l=3$ and $k=1$ and secret question $q=$``Is the picture somehow related to computation?''}
\label{fig8}
\end{figure}

In Section VI, we show that the total amount of work required by the adversary is: $\frac{{L^2  + 1}}{{l + 1}}\frac{n}{2}\log _2 n$. Putting in the values $L=10, l=5$ and $n=10^6$, we get a total work load of $ \approx 2^{27} $. Which is significantly higher than the previous protocol. The adversary thus gets the secret question. But how about the secret permutation string $\sigma $? The adversary is only successful in impersonating as the legitimate user if it knows $\sigma $. It turns out that after the secret question has been revealed, it just takes a handful of iterations to observe to guess the permutation string with high probability. We show this in Section V.
 
\paragraph*{Discussion on the protocol} Even though this protocol does have an advantage over the basic protocol in the sense that the correct answer sequence is shuffled, it seems hard to construct a method so as to hide the answer sequence completely with not too much effort on the user's part. It not only adds extra burden while answering the pictures, it also slows down the process. Secondly, in the two schemes proposed, there is a big question of practicality. 
How to automatically generate those pictures? We conjecture that it is hard to write a program that can extract all the features from the given features. But how about the problem of generating or finding out images satisfying a given question automatically? This might not be possible for all questions. In the next scheme we try to create a protocol that can automatically generate instances. 
\paragraph*{Matrix Interpretation} It would be interesting to know whether this protocol can be presented in a matrix representation. We could represent it as ${\bf{Vx}} = {\bf{a}}$. However, the actual answer sequence is not the same as ${\bf{a}}$ in this case. Thus we can represent it as: ${\bf{Vx}} = p_L \left( {\bf{a}} \right)$  where $p_L \left( {\bf{a}} \right)$ is a permutation operation on ${\bf{a}}$
 taking its $L$ components at a time. However, we still have the case that some of the bits in ${\bf{a}}$ are random. This becomes a problem similar to the Learning Parity with Noise (LPN) problem presented in the HB protocol \cite{hopper}. Again, the real problem in our scheme is to extract all features from the pictures and thus constructing a correct matrix ${\bf{V}}$.
\subsection{A Practical Scheme}
In this scheme, we would use the Gimpy CAPTCHA \cite{gimpy}. Gimpy works by picking several words from a dictuonary, distorting the text of these words and presenting the resulting words in the form of an image in front of a human user. The idea is that the current computer programs cannot comprehend the text. We assume a dictionary of size $N$. The algorithm $\mathsf{Gimpy}\left( j \right)$ does the same thing as gimpy, except that now it takes a desired number $j$, of words from the dictionary. Let $L$ be a small positive number, e.g. $11$. The whole image screen is divided into $L$ boxes. Let $s$ and $t$ be non-negative integers modulo $L$, kept as secret between the user and the computer. An initial value $x_0$, another non-negative number modulo $L$, is also kept as a secret. A secret question $q$ is constructed from the dictionary words. For example, $q$ could be: ``Are there more than three words begining with the letter ``B''?''. Let $\mathsf{Gimpy}\left( q, j \right)$  be the algorithm that takes $j$ pictures from the dictionary such that the resulting challenge satisfies the question $q$, and let $\mathsf{Gimpy}\left( -q, j \right)$ be the one that does not satisfy $q$. Let $\mathsf{Grid}\left( L \right)$ be the procedure that concatenates $L$ images (boxes) into one image in the form of a grid. The protocol is described as follows:\\\\
\textsc{Setup.} Randomly generate integers $s$, $t$ and $x_0$ modulo a public integer $L$, and share them as a secret between the server and the user. Share a secret question $q$ from ${\cal Q}$.\\
\textsc{Protocol.}
\begin{itemize}
\item For $i = 1$ to $k$, do:
	\begin{itemize}
	\item Compute $x_i  \equiv sx_{i - 1}  + t\bmod L$.
	\item Select a bit $b$ uniformly at random and apply $\mathsf{Gimpy}\left( q, j \right)$ to the $x_i$th box if $b=1$ and $\mathsf{Gimpy}\left( -q, j \right)$ otherwise. 
	\item For each of the remaining boxes, apply $\mathsf{Gimpy}\left( j \right)$. 
	\item Apply $\mathsf{Grid}\left( L \right)$, and present it as a challenge to the user.
	\item The user computes $x_i  \equiv sx_{i - 1}  + t\bmod L$ and submits $a = q\left( x_i \right)$, where $x_i$ denotes the $x_i$th box.
	\end{itemize}
\item Output $\mathsf{accept}$ if all answer's are correct, otherwise output $\mathsf{reject}$.
\end{itemize}
The protocol is described in the Figure 4. Here $L=4$, $s=3$, $t=3$, $x_0=5$. Thus $x_1=2$ and so the user looks at the picture labelled $2$ and answers the question $q=$``Does the picture contain the names of at least two animals?''. The pictures are taken from the Gimpy webpage \cite{gimpy}
\begin{figure}[ht]
\centerline{\includegraphics[width=4.63in,height=2.87in]{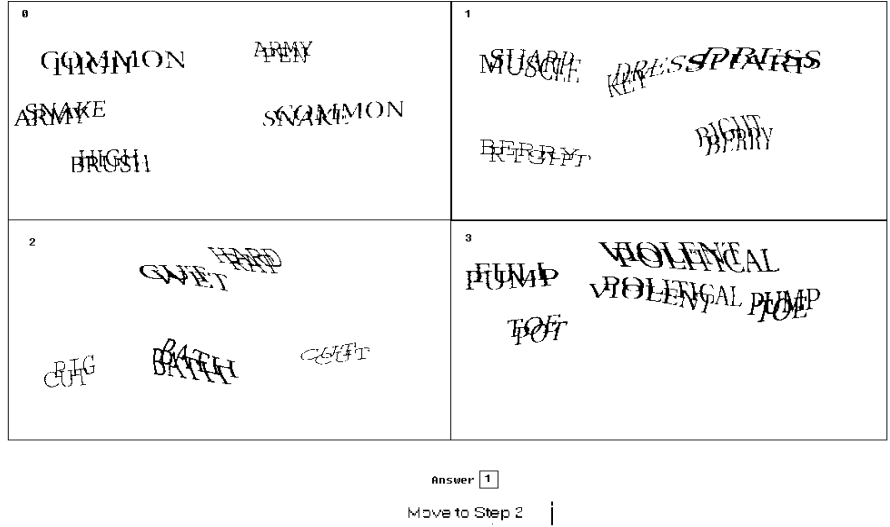}}
\caption{The Protocol Scheme with $L=4$, $s=3$, $t=3$, $x_0=5$ and secret question $q=$``Does the picture contain the names of at least two animals?''}
\label{fig7}
\end{figure}
How about the total amount of required by the adversary? We see that this protocol has the form of the previous protocol if we let $L=L$ and $l=1$. Thus the total amount of work required by the adversary is$\frac{{L^2  + 1}}{2}\frac{{n\left( N \right)}}{2}\log _2 \left( {n\left( N \right)} \right)$. The quantity $n\left( N \right)$ denotes that the total number of features (possible questions) are a function of the size $N$ of the dictionary. 
\paragraph*{Discussion on the protocol} The main theme of the protocol is to let the human adversary write down all the words presented in the image. Notice the use of the linear congruential generator $x_i  \equiv sx_{i - 1}  + t\bmod L$, for small values of $L$ and hence $s$ and $t$. This is used to induce randomness in the selection of the secret box. Although this is not a cryptographically strong pseudorandom number generator, and \textit{certainly not} for small values of $L$, the use is just there to inject some kind of randomness. Since the values of $x_i$ are not shown in the clear, it is safe enough for our purposes. Notice that for some values of the parameters $s$, $t$ and $x_0$, the $x_i$'s do not span the whole set of integers modulo $L$. This again is not a worry as the adversary does not know which parameters are chosen. But the most important concern is: How much do the distorted images of texts and numbers describe? Unfortunately, the full spectrum of features described by a distorted image of a word from a dictionary is not much more than a natural image. Therefore, we should conclude that although the above scheme puts some autonomy in the challenge generation process, it is not as secure as the ones which contain natural images in terms of the workload on the adversary. Details follow in Section VI.
\paragraph*{Matrix Interpretation} Once again, we are tempted to use a matrix representation for the protocol. We could represent it as $p\left( {\bf{V}} \right){\bf{x}} = {\bf{a}}$, where ${\bf{V}}$ represents the matrix of features of all the $L$ pictures in every iteration. And $p\left( {\bf{V}} \right)$ represents the function which picks one of those $L$ pictures according to chosen parameters of the linear congruential generator. Thus this protocol is opposite of the previous one in the sense that instead of diffusing the answer string, the pictures are sort of randomly chosen each time.  
\subsection{Using Multiple Questions}
In \cite{hassan} we proposed that we can have a group of questions as a secret connected by any combination of logical connectives, like AND, OR and NOT. However, we should not have greater than a certain number of logically connected features, because otherwise the workload on the legitimate user increases considerably. We could safely use a group of 3 or less questions. The adversary's algorithm described above and in the analysis will not work in this case, unless all the questions are connected by the logical AND. There are still ways to go around this, the adversary this time around looks for inconsistent features in the picture and eliminates them in each iteration. More precisely, the adversary's task is to find a boolean function consisting of 3 or less literals that satisfies the truth assignments of all the literals. More precisely, let ${\bf{V}}$ denote the matrix consisting of features extracted from the pictures. Each column of this matrix represents the absence or presence of a feature in the corresponding picture. If the adversary also knows the answer vector ${\bf{a}}$, then its job is to find a boolean function satisfying the mapping. We could use multiple questions in any three of the above protocols. The basic protocol is then reduced trivially to the problem of finding the boolean function defined above. The enhanced protocol however becomes a bit more tricky, since we do not know the exact evaluation of the expression. The practical scheme also becomes hard as we do not know the literals being used in the evaluation of the boolean function. 
There can be two variants of this protocol; The basic protocol and the enhanced protocol. Obviously, the adversary can eliminate inconsistent features in the basic protocol easily as it does not involve any random replies. However, if we use the enhanced protocol with questions connected by logical operators, we can make the adversary's task harder as there would be random bits in the answer sequence as well. 
\section{Experiments}
We did a few experiments in order to get an idea about the efficiency of our scheme. The experimental stage consisted of two main experiments: The first one was carried out to see how many distinct features can be extracted from a given picture; The second one was to check whether it is easy for a human user to tell whether a given feature is present in a picture or not. For the first experiment, we presented the image in Figure 1 to ten participants (all computer science graduate school students). Each one of them were asked to write down as many features as they believed the picture contained. A commulative total of 42 distinct features were extracted by the participants altogether, not counting the multiplicity of some features (such as the digits 1,2,...,9 are written together). These features are given in the Table 1 along with their frequency which means the number of participants who wrote down the corresponding feature:\\

\begin{table}
\caption{Experimental Results}

\begin{center} 
\begin{tabular}{|l|c|l|c|}
\hline
Feature&Freq \& Res&Feature&Freq \& Res\\
\hline
Numbers (Digits 1-9)&5 \tick\tick\tick&Digits without closed loop(s)&1 \tick\tick\tick\\
Black color&4 \tick\tick\tick&Digits with closed loop(s)&1 \tick\tick\tick\\
Columns and rows sum to 15&3 \tick\tick\tick&A heart&1 \tick\cross\tick\\
Diagonal sums to 15&3 \tick\tick\tick&Sign board&1 \tick\cross\cross\\
Square(s)&3 \tick\tick\tick&Triangle&1 \tick\cross\tick\\
Matrix&3 \tick\tick\tick&Cross(es)&1 \tick\cross\tick\\
3X3 matrix&3 \tick\tick\tick&Line(s)&1 \tick\tick\tick\\
White color&2 \tick\tick\tick&Rectangle(s)&1 \tick\tick\cross\\
Magic square&2 \tick\tick\tick&Stair(s)&1 \tick\tick\tick\\
Odd number(s)&2 \tick\tick\tick&'+'sign&1 \tick\tick\tick\\
Black line(s)&2 \tick\tick\tick&Circle(s)&1 \tick\tick\tick\\
Hook&1 \tick\cross\tick&Zig zag path&1 \tick\cross\tick\\
Slide&1 \tick\cross\tick&Alphabets C,S,L and O&1 \tick\cross\tick\\
'X' sign&1 \tick\cross\cross&The string 492357816&1 \tick\tick\tick\\
The string 438951276&1 \tick\tick\tick&Distortion in slanted line(s)&1 \tick\cross\tick\\
Table&1 \tick\cross\tick&Array&1 \tick\tick\tick\\
Balance&1 \cross\cross\cross&Equilibrium&1 \tick\cross\cross\\
Symmetry&1 \tick\tick\tick&Complements&1 \tick\tick\cross\\
Typed digit(s)&1 \tick\tick\tick&White area $>$ Black area&1 \tick\tick\tick\\
Even number(s)&1 \tick\tick\tick&The right angle&1 \tick\tick\tick\\
Mathematics&1 \tick\tick\cross&Arithmetic&1 \tick\tick\cross\\
\hline
\end{tabular}
\end{center} 
\end{table}

Once the features were collected, they were shown to three separate individuals not present in the first experiment. They were shown the picture and asked to answer whether the given list of features found by the participants in the first experiment were present in the picture or not. Their responses are shown in the column labeled "Freq and Res" where a \tick represents that the corresponding participant believed the feature to be present in the picture. Not surprisingly, the features with the higher frequencies were answered correctly by all three users. On the other hand, some of the single frequency features were also answered correctly by all three users. The ones with indifferent answers are those that require a `keen' eye, e.g. ``hook''.
These experiments show that even a simple picture as the one shown in Figure 1 can have a lot of features, majority of which are very easy to answer but not so easy to extract. There can still be a lot more features present in the picture; one such example is "`one side of a rubik's cube"'.
This survey gives us some guidelines while choosing the pictures and/or secret questions:
\begin{itemize}
	\item Do not use pictures whose main object is the secret feature. So for example, if we chose the picture of Figure 1 as the challenge picture, then the secret question of "`Does the picture contain the digits 1-9?"' will certainly be a bad choice.
	\item Do not use simple pictures. Simple pictures contain very few features. This is evident from Figure 1. Although, one may still be able to think of more features, there does not seem to be a big number of features.
	\item Do not use secret questions which are hard to answer by the legitimate user. As an example, the feature "`Equilibrium"' in the above table was answered "`no"' by two users. This seems hard to find out in the picture and needs more of a philosophical eye. 
	\item Always allow for user error. So for example, if the user replies '10' pictures, allow an error of 2 to 3 wrong answers. This is clear from the picture that a user answered "`no"' to the feature "`Mathematics"', even though it seems to be describe the figure. 
\end{itemize}
\section{Analysis of the Enhanced Protocol}
In this section, we would like to analyze the relationship between the hidden permutation $\sigma $ and the secret question $q$ in the enhanced protocol. The purpose of analysis is to find out the strength of the protocol if one of these secrets is leaked out. For the first part, let's assume that the adversary somehow found out the hidden permutation $\sigma$ but not the secret question. Since the adversary knows the hidden permutation, it knows exactly which questions are being answered and which one are being answered randomly. Thus the adversary can neglect the randomly answered pictures and use the pictures with correct answers to look for the secret question. Thus this transforms to the Basic Protocol in a straightforward manner.

For the other side, let us assume the adversary ${\cal H}$ who knows the secret question $q$. It evaluates each picture for upto $k$ iterations. Let $X_i \left( t \right)$ be the variable representing the evaluated bit of the $i$th picture in 
the $t$th iteration of the enhanced protocol, where $1 \le t \le k$. ${\cal H}$ thus evaluates the following information after $k$ iterations:
\[
\begin{array}{*{20}c}
   {X_1 \left( t \right)} & {X_2 \left( t \right)} &  \cdots  & {X_L \left( t \right)}  \\
   {b_{11} } & {b_{12} } &  \cdots  & {b_{1L} }  \\
   {b_{21} } & {b_{22} } &  \cdots  & {b_{2L} }  \\
    \vdots  &  \vdots  &  \vdots  &  \vdots   \\
   {b_{k1} } & {b_{k2} } &  \cdots  & {b_{kL} }  \\
\end{array}
\]
where each $b_{ij}$ is the bit evaluated by ${\cal H}$ for the corresponding picture. ${\cal H}$ also has the following response table from the legitimate user:
\[
\begin{array}{*{20}c}
   {Y_1 \left( t \right)} & {Y_2 \left( t \right)} &  \cdots  & {Y_L \left( t \right)}  \\
   {a_{11} } & {a_{12} } &  \cdots  & {a_{1L} }  \\
   {a_{21} } & {a_{22} } &  \cdots  & {a_{2L} }  \\
    \vdots  &  \vdots  &  \vdots  &  \vdots   \\
   {a_{k1} } & {a_{k2} } &  \cdots  & {a_{kL} }  \\
\end{array}
\]
where each $Y_i \left( t \right)$ represents the user's response bit to the $i$th picture in the $t$th iteration. The adversary now runs the following simple algorithm:

\begin{itemize}
\item Initialize $\sigma \left( . \right) = {\rm{null}}$.
\item For each $X_i \left( t \right)$, check whether there exists a $Y_j \left( t \right)$ such that the two match at every corresponding bit position.
	\begin{itemize}
	\item If there is only one such $Y_j \left( t \right)$ then mark $\sigma \left( j \right) = i$.
	\item If there are two such $Y_j \left( t \right)$'s then $\mathsf{halt}$.
	\end{itemize}
\item Assign * to each unassigned position of $\sigma \left( . \right)$ 
\item Output the permutation $\sigma$ and $\mathsf{halt}$.
\end{itemize}
We now state the following theorem:\\
\newtheorem{theo}{Theorem}
\begin{theo}
\[
\left( {\left( {1 - \frac{1}{{2^k }}} \right)^{L - 1} \left( {\frac{l}{L}} \right) + \left( {1 - \frac{1}{{2^k }}} \right)^L \left( {1 - \frac{l}{L}} \right)} \right)^L  \le \Pr \left[ {\sigma \,{\rm{is}}\,{\rm{correct}}} \right] \le \left( {1 - \frac{1}{{2^k }}\left( {1 - \frac{l}{L}} \right)} \right)^L 
\]

\end{theo}
\begin{figure}[ht]
\centerline{\includegraphics[width=3.63in,height=2.87in]{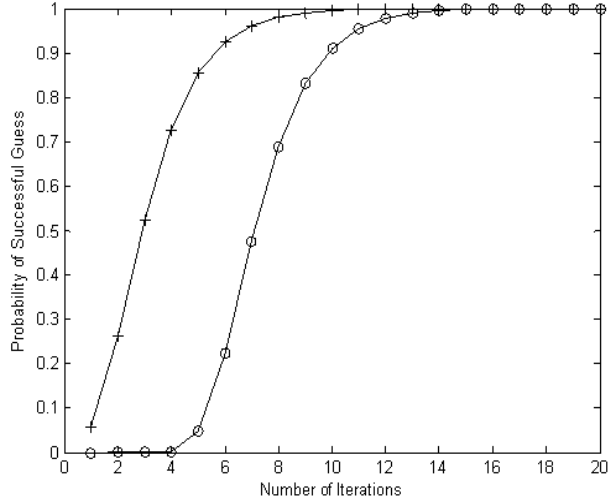}}
\caption{Guessing Probability}
\label{fig1}
\end{figure}

\begin{proof}
Let $A_i$ be the event that the adversary correctly guesses the $i$th position of the permutation $\sigma$. Without loss of generality, we assume that the adversary starts with the left most position and goes on to the next position in sequential order. We have to find the probability:
\[
\Pr \left[ {\sigma \,{\rm{is}}\,{\rm{correct}}} \right] = \Pr \left[ {A_1 } \right]\Pr \left[ {A_2 |A_1 } \right] \cdots \Pr \left[ {A_L |A_1  \wedge A_2  \wedge  \ldots  \wedge A_{L - 1} } \right]
\]
Now let $B_i$ be the event that position $i$ is not the don't care position. Let $\overline {B_i }$ be the complementary event. It is clear that:
\[
\Pr \left[ {A_i } \right] = \Pr \left[ {A_i |B_i } \right]\Pr \left[ {B_i } \right] + \Pr \left[ {A_i |\overline {B_i } } \right]\Pr \left[ {\overline {B_i } } \right]
\]
It is easy to see that: 
\[
\Pr \left[ {B_1 } \right] = \frac{l}{L}  \rm{ and }  \,\Pr \left[ {\overline {B_1 } } \right] = 1 - \frac{l}{L}
\]
If $B_1$ is true then $X_1 \left( t \right)$ matches at least one of the $Y_j \left( t \right)$'s . The adversary's algorithm will guess it correctly if there is only one such $Y_j \left( t \right)$. So:
\begin{eqnarray*}
  \Pr \left[ {A_1 |B_1 } \right]  &=&  \Pr \left[ {{\rm{There}}\,{\rm{exists}}\,{\rm{only}}\,{\rm{one}}\,j\,{\rm{such}}\,{\rm{that}}\,Y_j \left( t \right) = X_1 \left( t \right)} \right]\\  
    &=&  1 \times \underbrace {\left( {1 - \left( {\frac{1}{2}} \right)^k } \right) \times  \ldots  \times \left( {1 - \left( {\frac{1}{2}} \right)^k } \right)}_{L - 1\,{\rm{times}}} \\
    &=&  \left( {1 - \left( {\frac{1}{2}} \right)^k } \right)^{L - 1}  
\end{eqnarray*}

Now if $\overline {B_1 }$ is true, then the algorithm will detect this if $X_1 \left( t \right)$ does not match any of the $Y_j \left( t \right)$'s. The probability of this being true is:
\[
  \Pr \left[ {A_1 |\overline {B_1 } } \right] = \underbrace {\left( {1 - \left( {\frac{1}{2}} \right)^k } \right) \cdots \left( {1 - \left( {\frac{1}{2}} \right)^k } \right)}_{L\,{\rm{times}}} 
   = \left( {1 - \frac{1}{{2^k }}} \right)^L  
\]
With this we get:
\[
\Pr \left[ {A_1 } \right] = \frac{l}{L}\left( {1 - \frac{1}{{2^k }}} \right)^{L - 1}  + \left( {1 - \frac{l}{L}} \right)\left( {1 - \frac{1}{{2^k }}} \right)^L 
\]

Let us calculate the probability $\Pr \left[ {A_L |A_1  \wedge A_2  \wedge  \ldots  \wedge A_{L - 1} } \right]$. This is equal to:
\[
\begin{array}{l}
 \Pr \left[ {A_L |A_1  \wedge A_2  \wedge  \ldots  \wedge A_{L - 1} } \right] =\\
  \Pr \left[ {A_L |B_L  \wedge A_1  \wedge A_2  \wedge  \ldots  \wedge A_{L - 1} } \right]\Pr \left[ {B_L |A_1  \wedge A_2  \wedge  \ldots  \wedge A_{L - 1} } \right] + \\ 
 \Pr \left[ {A_L |\overline {B_L }  \wedge A_1  \wedge A_2  \wedge  \ldots  \wedge A_{L - 1} } \right]\Pr \left[ {\overline {B_L } |A_1  \wedge A_2  \wedge  \ldots  \wedge A_{L - 1} } \right] \\ 
 \end{array}
\]

It is straight forward to see that:\\
$\Pr \left[ {B_L |A_1  \wedge A_2  \wedge  \ldots  \wedge A_{L - 1} } \right] = \frac{l}{L}$ and $\Pr \left[ {\overline {B_L } |A_1  \wedge A_2  \wedge  \ldots  \wedge A_{L - 1} } \right] = 1 - \frac{l}{L}$. We also have:
\[
\Pr \left[ {A_L |B_L  \wedge A_1  \wedge A_2  \wedge  \ldots  \wedge A_{L - 1} } \right] = 1
\]
and
\[
\Pr \left[ {A_L |\overline {B_L }  \wedge A_1  \wedge A_2  \wedge  \ldots  \wedge A_{L - 1} } \right] = 1 - \frac{1}{{2^k }}
\]
We get the final result:
\[
\Pr \left[ {A_L |A_1  \wedge A_2  \wedge  \ldots  \wedge A_{L - 1} } \right] = \frac{l}{L} + \left( {1 - \frac{1}{{2^k }}} \right)\left( {1 - \frac{l}{L}} \right) = 1 - \frac{1}{{2^k }}\left( {1 - \frac{l}{L}} \right)
\]
Note that all other conditional probabilities for the events $A_2 ,A_3 , \ldots ,A_{L - 1}$ must lie between these two calculated probabilities. This gives us the upper and lower bounds for the probability of guessing the correct permutation as:
\[
\left( {\left( {1 - \frac{1}{{2^k }}} \right)^{L - 1} \left( {\frac{l}{L}} \right) + \left( {1 - \frac{1}{{2^k }}} \right)^L \left( {1 - \frac{l}{L}} \right)} \right)^L  \le \Pr \left[ {\sigma \,{\rm{is}}\,{\rm{correct}}} \right] \le \left( {1 - \frac{1}{{2^k }}\left( {1 - \frac{l}{L}} \right)} \right)^L 
\]\\
 
\end{proof}
\section{Obtaining the Secret Question}
Let $S = \left\{ {1,2, \ldots ,n} \right\}$ denote the universal set of all features. Let $A_1$ and $A_2$ denote two subsets of this set. In actual, $A_1$ and $A_2$, denote the set of features of two pictures drawn randomly according to an arbitrary distribution. We assume that any feature $i$ in $S$ is equally likely to occur in any of the subsets drawn. We define the following two indicator variables:
\[
\begin{array}{l}
 I_i  = \left\{ {\begin{array}{*{20}c}
   1 & {{\rm{if}}\,i \in A_1 }  \\
   0 & {{\rm{otherwise}}}  \\
\end{array}} \right. \\ 
 J_i  = \left\{ {\begin{array}{*{20}c}
   1 & {{\rm{if}}\,i \in A_2 }  \\
   0 & {{\rm{otherwise}}}  \\
\end{array}} \right. \\ 
 \end{array}
\]
Then we have that, $\left| {A_1  \cap A_2 } \right| = \sum\limits_{i = 1}^n {I_i J_i }$. The number of subsets of $S$ containing a given feature $i$ would be $2^{n - 1}$. Since each subset of $S$ is equally likely to contain $i$ regardless of the distribution with which the subset is drawn out, we have:
\[
\Pr \left[ {A_1 \,{\rm{contains}}\,i} \right] = \Pr \left[ {A_2 \,{\rm{contains}}\,i} \right] = {{2^{n - 1} } \mathord{\left/
 {\vphantom {{2^{n - 1} } {2^n }}} \right.
 \kern-\nulldelimiterspace} {2^n }} = {1 \mathord{\left/
 {\vphantom {1 2}} \right.
 \kern-\nulldelimiterspace} 2}
\]
From this and the fact that the two subsets are drawn independently of each other, we have:
\[
E\left[ {\left| {A_1  \cap A_2 } \right|} \right] = \sum\limits_{i = 1}^n {E\left[ {I_i J_i } \right]}  = \sum\limits_{i = 1}^n {E\left[ {I_i } \right]} E\left[ {J_i } \right] 
\]
Now, $E\left[ {I_i } \right] = \sum\limits_{A_1  \subseteq S} {I_i \Pr \left[ {A_1 \,{\rm{is}}\,{\rm{chosen}}\,{\rm{from}}\,S} \right]}  = {{2^{n - 1} } \mathord{\left/
 {\vphantom {{2^{n - 1} } {2^n }}} \right.
 \kern-\nulldelimiterspace} {2^n }} = {1 \mathord{\left/
 {\vphantom {1 2}} \right.
 \kern-\nulldelimiterspace} 2}$. Similarly, $E\left[ {J_i } \right] = {1 \mathord{\left/
 {\vphantom {1 2}} \right.
 \kern-\nulldelimiterspace} 2}$. Finally, this gives us:
\[
E\left[ {\left| {A_1  \cap A_2 } \right|} \right] = \sum\limits_{i = 1}^n {E\left[ {I_i } \right]} E\left[ {J_i } \right] = {n \mathord{\left/
 {\vphantom {n 4}} \right.
 \kern-\nulldelimiterspace} 4}
\]

In general we can see that: 
\[
E\left[ {\left| {A_1  \cap A_2  \cap  \ldots  \cap A_t } \right|} \right] = \frac{n}{{2^{t + 1} }} = \frac{1}{2}E\left[ {\left| {A_1  \cap A_2  \cap  \ldots  \cap A_{t - 1} } \right|} \right]
\]

We analyze the three protocols using the result obtained above. According to over discussion in Section III, the adversary will try to narrow down the number of possible secret features by using the algorithm $\mathsf{GetBasicQ}$ described in Section III. 
 
\subsection{The Basic Protocol}
The adversary ${\cal H}$ looks at the current picture and its answer given by the user. It performs the procedure called $\mathsf{GetBasicQ}$ in described in Algorithm 1. Notice that ``Compute $A_k$'' means extracting the features of picture $A_k$.

\algsetup{indent=2em}
\begin{algorithm}[h!]
\caption{$\mathsf{GetBasicQ}$}\label{alg:factorial}
\begin{algorithmic}[1]
\REQUIRE A set of pictures $A_1 ,A_2 , \ldots$ together with their answers $a_1 ,a_2 , \ldots $
\ENSURE The secret feature $q$
\medskip
\IF {the answer bit position is $0$}
	\STATE Wait for the next iteration.
\ELSE 
	\STATE Extract the features in the picture as $A_1$
	\REPEAT
		\STATE For each picture $k$:
		\IF{$a_k  = 1$ (The answer bit at position $j$}
			 \STATE Compute $A_k$ (The picture at position $i$) and assign $A_1  \leftarrow A_1  \cap A_k $.
		\ELSE
			 \STATE As $a_k  = 0$ (The answer bit at position $j$), so compute $A_k$ (The picture at position $i$) and assign $A_1  \leftarrow A_1  - A_k $. 
		\ENDIF
	\UNTIL $\left| {A_1 } \right| = 1$ and $\mathsf{halt}$.
\ENDIF
\end{algorithmic}
\end{algorithm}


Next we compute the expected number of steps the adversary has to wait until he gets the above algorithm to halt.

\newtheorem{theo2}{Theorem}
\begin{theo2}
If ${\cal A}$ successfully extracts all the features, then the expected number of steps is $\log _2 n$
\end{theo2}

\begin{proof}
Let ${A_1^{\left( k \right)} }$ be the set ${A_1 }$ after the $k{\rm{th}}$ step. Our inductive proof is as follows:
First assume that $k=2$. There are two cases: First if  $a_k  = 1$ then as computed above:
\[
E\left[ {A_1^{\left( 2 \right)} } \right] = E\left[ {\left| {A_1^{\left( 1 \right)}  \cap A_2 } \right|} \right] = {n \mathord{\left/
 {\vphantom {n {2^2 }}} \right.
 \kern-\nulldelimiterspace} {2^2 }}
\]
And, if $a_k  = 0$, we also get the result:
\[
E\left[ {A_1^{\left( 2 \right)} } \right] = E\left[ {\left| {A_1^{\left( 1 \right)}  - A_2 } \right|} \right] = E\left[ {\left| {A_1^{\left( 1 \right)}  \cap A_2^c } \right|} \right] = {n \mathord{\left/
 {\vphantom {n {2^2 }}} \right.
 \kern-\nulldelimiterspace} {2^2 }}
\]

This is true, since ${A_2^c }$, the complement of ${A_2 }$, is also a subset of $S$. 

Now, in general for $k=t$, we have:
\[
E\left[ {A_1^{\left( t \right)} } \right] = {n \mathord{\left/
 {\vphantom {n {2^t }}} \right.
 \kern-\nulldelimiterspace} {2^t }}
\]
So, if $a_t  = 1$, then:
\[
E\left[ {A_1^{\left( {t + 1} \right)} } \right] = E\left[ {\left| {A_1^{\left( t \right)}  \cap A_{t + 1} } \right|} \right] = \frac{1}{2}E\left[ {A_1^{\left( t \right)} } \right] = \frac{n}{{2^{t + 1} }}
\]
And, if $a_t  = 0$, then again:
\[
E\left[ {A_1^{\left( {t + 1} \right)} } \right] = E\left[ {\left| {A_1^{\left( t \right)}  - A_{t + 1} } \right|} \right] = E\left[ {\left| {A_1^{\left( t \right)}  \cap A_{t + 1}^c } \right|} \right] = \frac{1}{2}E\left[ {A_1^{\left( t \right)} } \right] = \frac{n}{{2^{t + 1} }}
\]

The adversary will stop for some $k$ if $E\left[ {A_1^{\left( k \right)} } \right] = 1$, this means that:
\[
{n \mathord{\left/
 {\vphantom {n {2^k }}} \right.
 \kern-\nulldelimiterspace} {2^k }} = 1 \Rightarrow k = \log _2 n
\]
\end{proof}

Now, the number of iterations (pictures) for the adversary to observe are $\log _2 n$. At each step the adversary has to extract the features of a picture, hence the expected amount of work at each step is ${n \mathord{\left/
 {\vphantom {n 2}} \right.
 \kern-\nulldelimiterspace} 2}$
 and hence the total amount of work to be done by the adversary in the basic protocol is: $\frac{n}{2}\log _2 n$. How about the probability of success of the algorithm? We have assumed in this analysis that the adversary can extract all features in the image. In general, an adversary might not be able to extract everything in an image, including the secret feature. We can associate an average probability of $p$ with the extraction secret feature, which shows that the secret will be extracted at an average probability of $p$ whenever the adversary is presented with a picture with answer '1'. The '1' instances occur with an equal probability of ${1 \mathord{\left/
 {\vphantom {1 2}} \right.
 \kern-\nulldelimiterspace} 2}$. Therefore, the average probability of success of the above algorithm is: $p^{\frac{{\log _2 n}}{2}} $. In the special case, where the probability of the secret picture being extracted out is ${1 \mathord{\left/
 {\vphantom {1 2}} \right.
 \kern-\nulldelimiterspace} 2}$, the average probability would be: $\left( {\frac{1}{2}} \right)^{\frac{{\log _2 n}}{2}}  = \frac{1}{{\sqrt n }}$.

\subsection{The Enhanced Protocol}
Now suppose the adversary ${\cal H}$ wants to find out the hidden question in the enhanced protocol. This time the adversary cannot use the simple procedure $\mathsf{GetBasicQ}$ it used for the basic protocol because of the use of the permutation $\sigma$. It has to be selective in its choices. This time the adversary has to use a slightly modified version of $\mathsf{GetBasicQ}$ called $\mathsf{GetEnhancedQ}$ described in Algorithm 2. 


\algsetup{indent=2em}
\begin{algorithm}[h!]
\caption{$\mathsf{GetEnhancedQ}$}\label{alg:facty}
\begin{algorithmic}[1]
\REQUIRE A $ \ge L$ set of pictures $A_1 ,A_2 , \ldots$ together with an equal number of bits $a_1 ,a_2 , \ldots $
\ENSURE The secret feature $q$
\medskip
\STATE Select a random picture position $i$ between $1$ and $L$.
\STATE Select a random answer position $j$ between $1$ and $L$.
\IF {the answer bit position is $0$}
	\STATE Wait for the next iteration.
\ELSE 
	\STATE Extract the features in the picture as $A_1$
	\REPEAT
		\STATE For each iteration $k$:
		\IF{$a_k  = 1$}
			 \STATE Compute $A_k$ and assign $A_1  \leftarrow A_1  \cap A_k $.
		\ELSE
			 \STATE As $a_k  = 0$, so compute $A_k$ and assign $A_1  \leftarrow A_1  - A_k $. 
		\ENDIF
	\UNTIL $\left| {A_1 } \right| = 1$ and $\mathsf{halt}$.
\ENDIF
\end{algorithmic}
\end{algorithm}


Now assume that the adversary does the following: Whenever it has guessed an incorrect path, it executes the above algorithm for an expected number of $\log _2 n$ steps and then goes back again to choose a different path (This shows the expected time until the adversary realizes that it has chosen the wrong picture and answer pair). For each picture position $1 \le i \le L$, there are a possible $L$ answer positions. Each of these are equally likely for the adversary to pick. Out of these, only $l$ result in halting the algorithm. If the adversary, found the correct path, it will stop by outputting the feature. This can occur with probability ${l \mathord{\left/
 {\vphantom {l {L^2 }}} \right.
 \kern-\nulldelimiterspace} {L^2 }}$. If the adversary chooses the wrong path, it will go back again and choose another path. The total number of correct paths in the second iteration would be: $l\left( {L^2  - 1} \right)$. Thus the probability of the adversary stopping after two iterations would be:$\frac{l}{{L^2 }}\left( {\frac{{L^2  - l}}{{L^2  - 1}}} \right)$. Continuing in this fashion, if we let $\Pr \left[ {y_i } \right]$ denote the probability of the adversary stopping at the $i$th step, we get:
\[
\Pr \left[ {y_i } \right] = \frac{l}{{L^2 }}\left( {\frac{{L^2  - l}}{{L^2  - 1}}} \right) \cdots \left( {\frac{{L^2  - l - \left( {i - 2} \right)}}{{L^2  - \left( {i - 1} \right)}}} \right),\,\,\,\,\,\,\,\,\,\,{\rm{for}}\,i \ge 2
\]

Each of these paths results in the termination of the algorithm after a certain number of steps by finally choosing the correct path. We can now find the expected number of steps of the adversary:
 
Let $Y$ denote the number of steps taken by the adversary. Thus $y_i  \in Y$ denotes the number of steps in the $i$th path. We get:
\begin{eqnarray*}
  E\left[ Y \right] &=& \frac{l}{{L^2 }}\log _2 n + \frac{l}{{L^2 }}\left( {\frac{{L^2  - l}}{{L^2  - 1}}} \right)\left( {2\log _2 n} \right) +  \cdots  + \frac{l}{{L^2 }}\left( {\frac{{L^2  - l}}{{L^2  - 1}}} \right)\left( {\frac{{L^2  - l - 1}}{{L^2  - 2}}} \right)\\
  &\cdots& \left( {\frac{{L^2  - l - \left( {L^2  - l - 1} \right)}}{{L^2  - \left( {L^2  - l} \right)}}} \right)\left( {L^2  - l + 1} \right)\log _2 n\\  
   &=& \frac{l}{{L^2 }}\log _2 n( 1 + 2\left( {\frac{{L^2  - l}}{{L^2  - 1}}} \right) +  \cdots  + \left( {L^2  - l + 1} \right)\left( {\frac{{L^2  - l}}{{L^2  - 1}}} \right)\left( {\frac{{L^2  - l - 1}}{{L^2  - 2}}} \right)\\
   &\cdots& \left( {\frac{{L^2  - l - \left( {L^2  - l - 1} \right)}}{{L^2  - \left( {L^2  - l} \right)}}} \right) ) \\ 
   &=& \frac{l}{{L^2 }}\log _2 n\frac{{\left( {L^2  - l} \right)!}}{{\left( {L^2  - 1} \right)!}}\sum\nolimits_{i = 1}^{L^2  - l + 1} {\left( {i\frac{{\left( {L^2  - i} \right)!}}{{\left( {L^2  - l - i + 1} \right)!}}} \right)} \\ 
   &=& \frac{l}{{L^2 }}\log _2 n\frac{{\left( {L^2  - l} \right)!}}{{\left( {L^2  - 1} \right)!}}\frac{{L^2 \left( {L^2  + 1} \right)\left( {L^2  - 1} \right)!}}{{l\left( {l + 1} \right)\left( {L^2  - l} \right)!}} \\
   &=& \frac{{L^2  + 1}}{{l + 1}}\log _2 n 
\end{eqnarray*}
In light of the previous reult, the total amount of work done by the adversary is $\frac{{L^2  + 1}}{{l + 1}}\frac{n}{2}\log _2 n$. The probability of success of the algorithm depends on the whether the adversary has chosen the correct combination of picture and answer position pair. Thus if we again let $p$ be the average proabibility of successfully extracting the feature, the result come out to be the same as before: $p^{\frac{{\log _2 n}}{2}} $.
\subsection{The Practical Scheme}
We could analyze the workload in the practical scheme by viewing the behavior of the linear congruential generator for small values of the arguments. However, for simplicity, we can assume that the position of the next picture is determined analogous to the previous protocol. Thus we can let $L=L$ and $l=1$ in our result for the previous protocol. What about the value of $n$? Ofcourse this should depend on the dictionary size $N$ which can be anywhere in the range of $10^3$ to $10^5$. However, $n$ represents the number of distinct features and this could not be possibly more than $N$. We say this because two words might contain the same letters like ``wolf'' and ``flow'', and two words might represent the same concept, like synonyms. Therefore we can assume $n = xN$ where $0 < x < 1$. Assuming $x=0.5$ we get the following result for adversary's work: $\frac{{p^2  + 1}}{2}\frac{N}{4}\log _2 \left( {\frac{N}{2}} \right)$. The probability of success in this case would also come out to be $p^{\frac{{\log _2 n}}{2}} $.
\subsection{Comparative Workloads of the Three Protocols}
Based on the results obtained in the previous subsections, we can show the comparative workloads on the adversary in the three protocols. First we show the workload by fixing $n = 10^5 $ and $N = 10^4 $ and plotting the three graphs as a function of $L$. In the enhanced scheme we have assumed $l = \left\lceil {\frac{L}{2}} \right\rceil $. The three plots are shown in Figure 6.
\begin{figure}[ht]
\centerline{\includegraphics[width=3.63in,height=2.87in]{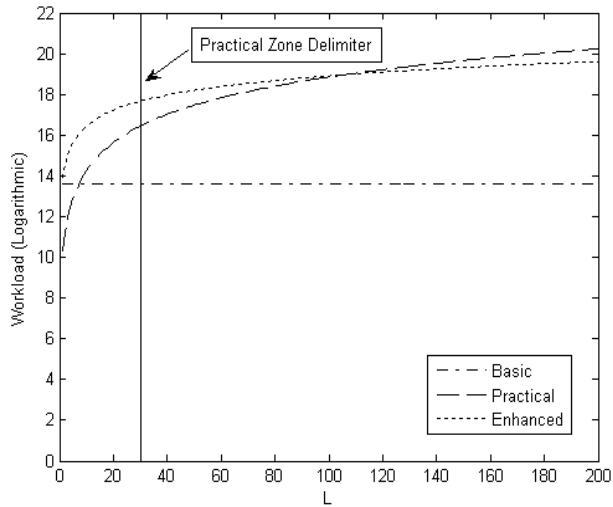}}
\caption{Comparative workloads of the adversary in the three schemes}
\label{fig2}
\end{figure}

%
\begin{figure}[ht]
\centerline{\includegraphics[width=3.63in,height=2.87in]{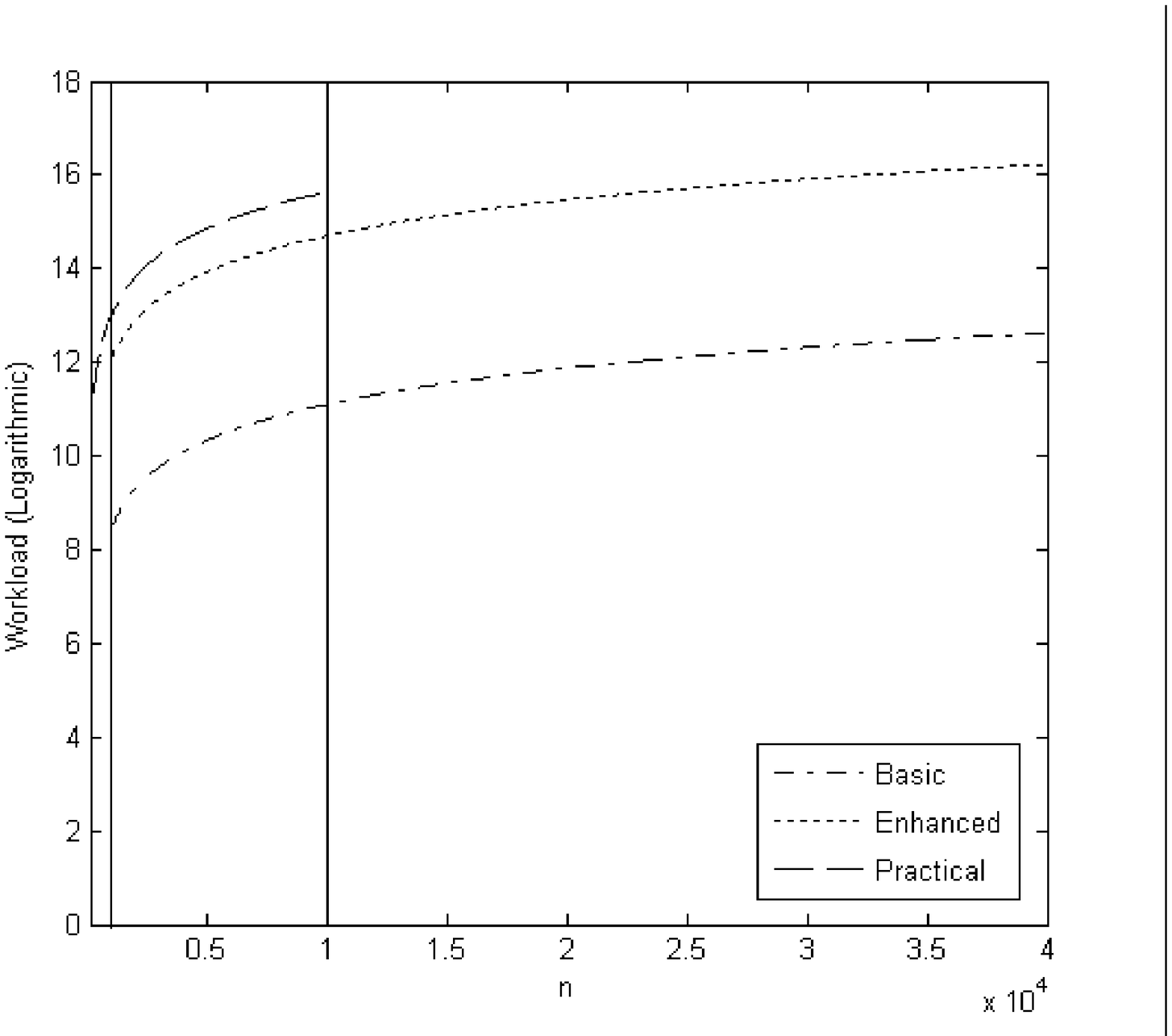}}
\caption{Comparative workloads of the adversary in the three schemes with changing n}
\label{fig3}
\end{figure}

The workload of the adversary increases in the two schemes with the increasing value of $L$ as compared to the basic protocol. Notice the use of ``Practical Zone Delimiter''. This is placed at a value of $L=25$, since we believe that putting more pictures in a given iteration would indeed place computational burden on the human user. For small values of $L$, we see that the Basic and the Enhanced schemes work better as compared to the Practical Scheme. As we increase $L$ even beyond the practical zone, the practical scheme becomes better. But this is because we have fixed $l$ to be one half $L$ in the enhanced scheme. The advantage of the two schemes over the basic scheme does not come without a disadvantage. The memory and processing requirements of the other two schemes also increase with an increment in $L$. The following table shows the comparison:

\begin{table}
\caption{Qualitative comparison of the three schemes}
\begin{center}

\begin{tabular}{|l|c|l|c|}
\hline
Scheme&Image Evaluation&Memory&Computation\\
\hline
Basic& \tick & $q$ & \cross\\
Enhanced& \tick & $q$ and $\sigma \left( {L,l} \right)$& \cross\\
Practical& \tick & $q$ and $a,b,x_i$ all $ < L$& 1 multiplication and \\ &&& addition modulo $L$\\
\hline
\end{tabular}

\end{center}
\end{table}

Figure 7, shows the comparative workloads as a function of $n$ ($N$ in the practical scheme case). The value of $L$ is fixed at 20. The range of $n$ is from $10^3$ to $10^5$ and that of $N$ is from $10^2$ to $10^4$. Interestingly, the enhanced scheme becomes better with larger values of $n$ as should be evident from the fact that $N$ has a much smaller value as compared to $n$.

Finally we show the interrelationship between $l$ and $L$ in the enhanced shceme. We see that the workload of the scheme increases significantly with lower values of $l$ and higher values of $L$. This is shown in Figure 8.
\begin{figure}[ht]
\centerline{\includegraphics[width=3.63in,height=2.87in]{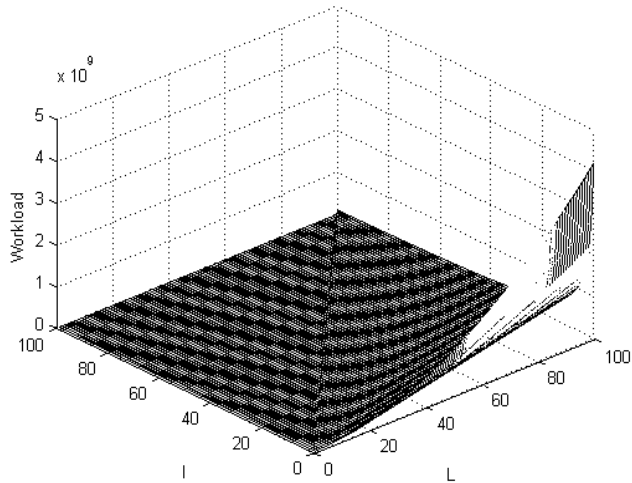}}
\caption{Graph of the adversary's workload in the enhanced scheme with changing $l$ and fixed $L$}
\label{fig4}
\end{figure}
\section{Conclusion and Future Work}
Human identification protocols can be a good alternative for the traditionally less secure password based systems. Over the years researchers have tried to construct efficient Human identification protocols which are secure against passive or active adversaries. However, the protocols run short in terms of efficiency and security. One such protocol was proposed by us in \cite{hassan} and its security was based on the ``conjectured difficulty'' of obtaining the secret after observing some authentication sessions. In this paper, we have extended the work by giving some candidate alternative protocols, finding the exact harndess of these protocols in terms of the effort required by the human adversary as well as giving a detailed analysis of the protocol proposed in \cite{hassan}. A brief survey regarding the number of possible features in an image was also carried out. Our results show that a practical implementation of the protocol might be feasile provided we are against a resource constrained human adversary.

A notable future line of work is to device a similar protocol secure against active adversaries. This might involve sending challenges to the server by the human user similar to \cite{pervasive}. However, it remains an open problem whether we can fine tune the protocols so as to make them secure against active adversaries without increasing the workload on legitimate users. Another direction of future work is to come up with a different model for the distribution of features in images found on the web. This might give a close to realistic quantification of the workload on the adversary.

\end{document}